# High accuracy, high dynamic range optomechanical accelerometry enabled by dual comb spectroscopy

D. A. Long,[1] J. R. Stroud,[2] B. J. Reschovsky,[1] Y. Bao,[1] F. Zhou,[1] S. M. Bresler,[3] T. W. LeBrun,[1] D. F. Plusquellic,[2] J. J. Gorman[1]

[1]National Institute of Standards and Technology, Gaithersburg, Maryland, USA

[2]National Institute of Standards and Technology, Boulder, Colorado, USA

[3]University of Maryland, College Park, Maryland, USA

**Abstract**

Cavity optomechanical sensors can offer exceptional sensitivity; however, interrogating the cavity motion with high accuracy and dynamic range has proven to be challenging. Here we employ a dual optical frequency comb spectrometer to readout a microfabricated cavity optomechanical accelerometer, allowing for rapid simultaneous measurements of the cavity's displacement, finesse, and coupling at accelerations up to 24 g (236 m/s$^2$). With this approach, we have achieved a displacement sensitivity of 2 fm/Hz$^{1/2}$, a measurement rate of 100 kHz, and a dynamic range of $7.6 \times 10^5$ which is the highest we are aware of for a microfabricated cavity optomechanical sensor. In addition, comparisons of our optomechanical sensor coupled directly to a commercial reference accelerometer show agreement at the 0.5% level, a value which is limited by the reference's reported uncertainty. Further, the methods described herein are not limited to accelerometry but rather can be readily applied to nearly any optomechanical sensor where the combination of high speed, dynamic range, and sensitivity is expected to be enabling.

Cavity optomechanics provides significant advantages over conventional sensor transduction methods, such as capacitive and piezoelectric mechanisms, including higher sensitivity and bandwidth, as well as measurement traceability linked to optical frequencies. Optomechanical sensors have been successfully developed for a number of different physical measurements, including force[1, 2], pressure[3, 4], acceleration[5, 6], and ultrasound[7, 8]. While the benefits of optomechanical sensors are clear, one challenge is that an optical resonance frequency must be measured to determine changes in the cavity's length, which can then be converted to a particular measurement of interest, such as acceleration or force. This is most often done by locking a laser to the cavity and measuring the changes in the laser transmission or reflection to yield the effective length change. This approach can be effective when the frequency excursions are small but can severely limit the dynamic range and measurement rate because of the restricted locking bandwidth and tuning range of many lasers[9-11].

An alternative approach for measuring length changes of a cavity uses optical frequency combs to interrogate a given optical resonance[12-14]. With this approach, an electro-optic (EO) frequency comb is generated which can directly interrogate the sensor without the need for laser-to-cavity locking. This method allows for a robust and high dynamic range measurement of not only the resonance frequency but also the optical coupling efficiency and cavity finesse.

Here, we report on the use of a dual EO frequency comb spectrometer to perform rapid readout of a microfabricated optomechanical accelerometer. In our previous dual comb work,[13] we were limited in the dynamic range of this method by the onset of extended oscillations in the frequency domain cavity readout data. As we show here in model calculations, the cavity oscillations arise from magnification of rapid passage (dynamic response) effects from application of our fast dual frequency chirp approach. While this magnified view serves to give a sensitive measure of the cavity finesse, the extended frequency domain responses degrade the rapid measurements needed for precise cavity position sensing. To resolve this problem, we introduce both quadratic and linear phase terms to the inverse Fourier transformation which returns the data to the time domain and eliminates the rapid passage effects. This allows for high-fidelity and high-speed measurements of the cavity motion. With this approach, we have achieved, to the best of our knowledge, the highest dynamic range measurement of a microfabricated optomechanical sensor. In addition, we compare the accuracy of the accelerometer to a commercial reference accelerometer where both devices are stacked and simultaneously vibrated by the same shaker table. Below 1 kHz, the two devices agree to within the 0.5 % uncertainty reported for the calibrated reference standard.

The microfabricated optomechanical accelerometer has a planoconcave Fabry-Pérot geometry that is composed of a silicon concave micromirror with a radius of curvature of 273 µm and a planar mirror on a 11.1 mg proof mass that is suspended on silicon nitride beams to allow for transduction of acceleration into cavity length displacement[6, 14]. The optical cavity has a free-spectral range of 740 GHz (length of 203 µm) and a finesse of ≈ 5200 at the operating wavelength of 1588 nm. The mechanical quality factor and resonance frequency were determined through a thermomechanical noise measurement[6, 14] to be 115(1) and 23.686(1) kHz, respectively, where the uncertainties (Type A, k=1) were obtained directly from the fits. Increasing the mechanical resonance frequency, $\omega_0$, relative to previous devices[6, 14] has allowed for the measurement of larger accelerations, $a_e$, for a given cavity displacement, $x$ (i.e., $x \approx a_e/\omega_0^2$ for frequencies near DC). The accelerometer was coupled to a polarization-maintaining fiber and packaged in a fabricated stainless-steel enclosure to allow for mounting on a commercial electromechanical shaker table (see Fig. 1a). The shaker table employed for the present measurements was equipped with a reference accelerometer that provided a well-known acceleration with a standard uncertainty of 0.5% using closed loop operation.

For the largest accelerations demonstrated herein (> 230 m/s²), the cavity resonance was traveling over 240 cavity linewidths on short time scales. These large displacements generally preclude the use of traditional laser locking approaches but are well suited to interrogation by EO frequency combs. This method, which has been described previously[12, 13], utilizes an EO frequency comb to rapidly measure the cavity resonance mode shape and position with an acquisition time of 1 μs to 10 μs. A schematic of the EO dual comb spectrometer employed in the present measurements is shown in Fig. 1b. The primary difference between this instrument and the one described previously[13] is the use of balanced detection. Two frequency combs were generated using electro-optic phase modulators driven with linear chirped pulse waveforms produced by a fast arbitrary waveform generator. One of the frequency combs (referred to as the signal, SIG) interrogated the optomechanical accelerometer, while the second served as a local oscillator (LO). The chirped pulse applied to the SIG leg spanned $f_{start}$ = 0.1 GHz to $f_{stop}$ = 11 GHz. Simultaneously, the LO was chirped with a slightly different bandwidth (by ±400 MHz) but with the same $\tau_{CP}$ = 1 μs duration. After mixing on the photodetector, the LO chirp down-converts the SIG into a radiofrequency (RF) domain spectrum that spans $f_{start} \pm RF_{start}$ to $f_{stop} \pm RF_{stop}$, where $\Delta f_{RF} = RF_{stop} - RF_{start}$ = 400 MHz is the RF chirp bandwidth. Figure 2a shows how the SIG chirp and LO chirp relate in frequency over the duration of one chirp. The comb lines of the positive and negative EO sidebands are separated by the frequency difference between the pair of acousto-optic modulators (AOMs), while the comb lines arising from higher order EO sidebands were separated by applying a linear phase shift over the $N_{chirps}$ waveforms[15, 16].

For each comb measurement, the data from the down-converted interferogram was recorded and subsequently divided into 10 μs long sub-interferograms, each of which consisted of 10 repeated chirps (i.e., $N_{chirps}$ = 10). These sub-interferograms were then Fourier transformed to produce a frequency comb spectrum, referred to here as $RF_{RES}$. We normalized these spectra against a background comb spectrum, $RF_{BKG}$, which was acquired when the laser was detuned from any cavity resonance.

While similar comb generation through single-frequency drive methods can produce dual combs with the desired optical bandwidth and resolution, there is a tradeoff with the acquisition speed. In this case, it is well known for the dual comb readout that the RF bandwidth is limited to half of the optical comb tooth spacing[17]. This results in the definition of the acquisition speed limit as,

$$\Delta t_{min} = \frac{2\Delta f_{SIG}}{f_r^2} \tag{1}$$

where the $\Delta f_{SIG}$ is the optical bandwidth, and $f_r$ is the comb tooth spacing. To compare this to our chirped pulse system, we substitute an optical resolution of ≈ 27 MHz and the optical bandwidth of a single sideband of 10.9 GHz. For these parameters, Eq 1 gives a theoretical limit of approximately 30 μs for the traditional dual comb acquisition speed. This is three times larger than the 10 μs record length used for the chirped pulse system with a value of $N_{chirps}$ = 10. We also note the value of $N_{chirps}$ allows for the interleaving of 10 higher order combs and can be reduced to proportionally increase the acquisition speed since only the first-order comb was used here. The readily generated flat frequency combs from a single modulator in combination with the higher acquisition speed for a given optical resolution make the chirped approach more desirable for this and other potential applications.

As in our previous study[13], we observed rapid passage effects in these normalized comb spectra arising from the fast optical frequency chirps (see the Supplemental Material for further discussion). This leads to a distorted cavity line shape such as that shown in Fig. 2c as well as a reduced displacement sensitivity due to difficulties in spectral line shape fitting.

In the present work, we have applied a method to effectively remove the observed oscillatory response (see Fig. 2c) based on the inverse Fourier transform and the quadratic phase relation between the frequency and time domain spectra[15,16]. In general, we define the quadratic phase as:

$$\phi_i(t^2) = \pi \frac{\Delta f_i}{\tau_{CP}} t^2, \quad (2)$$

where $\Delta f_i$ is the chirp bandwidth and the subscript $i$ denotes either SIG, LO, or RF. The frequency domain phase terms, $\Phi_i$, are defined by the same time and frequency domain constants. The quadratic phase terms of the RF comb, $\phi_{RF}$, without any sample or device can be described by the phases, $\phi_{SIG}$ and $\phi_{LO}$, as,

$$RF_{BKG}(t) = A_{BKG}(t)\exp(i\phi_{RF}(t^2)) = A_{LO}(t)A_{SIG}(t)\exp\left(i(\phi_{LO}(t^2) - \phi_{SIG}(t^2))\right) \quad (3)$$

where $A_{SIG}$, $A_{BKG}$ and $A_{LO}$ are the magnitudes of the signal (SIG), down converted radio-frequency background ($RF_{BKG}$) and LO waveforms. However, the transient response from the cavity resonance will acquire the full quadratic phase from the LO comb. This can be described by the transient response, $A_T(t)$, mixed with the LO as,

$$T(t) = A_T(t)A_{LO}(t)\exp(i\phi_{LO}(t^2)) \quad (4)$$

The total system response when a resonance is present, $RF_{RES}(t)$, will be,

$$RF_{RES}(t) = A_{LO}(t)[A_{SIG}(t)\exp(i\phi_{RF}(t^2)) + A_T(t)\exp(i\phi_{LO}(t^2))] \quad (5)$$

Figure 2b shows how the quadratic phases of the SIG and LO mix to down-convert the spectra into the RF domain, while the transient response is magnified by acquiring the quadratic phase of the LO. Because the quadratic phase defines how the signal is translated into the frequency domain, the transient response of the cavity resonance gets magnified by the ratio between the LO phase, $\phi_{LO}$, and the RF phase, $\phi_{RF}$. The magnification factor is defined as:

$$m = \frac{\phi_{LO}}{\phi_{RF}}. \quad (6)$$

For the present experimental conditions, this magnification factor was 26.25 for $\Delta f_{SIG} < \Delta f_{LO}$ and 28.25 for $\Delta f_{SIG} > \Delta f_{LO}$. This magnified transient response results in the asymmetric rippled line shape seen in the frequency domain spectrum in Fig. 2c that is in excellent agreement with the modeled dynamic responses (see Section SIII).

Here we largely mitigate these rapid passage effects by converting the data back into the time domain as[16]:

$$RF_n(t) = \text{FT}^{-1}\{RF_n(\omega)\}, \quad (7)$$

where $RF_n(\omega)$ are the complex sampled frequency domain comb teeth, $n = 1\ldots400$ for each sideband, and $FT^{-1}\{\}$ is the inverse Fourier transform operator. This process dramatically reduces the distortion of the optical cavity resonance. However, some residual distortion in the time domain remains that introduces smaller errors in the retrieved acceleration (see discussion and Fig. S1 in Supplemental Materials Section II). These errors depended on the relative bandwidths, or magnification, of chirped waveforms: for $\Delta f_{SIG} > \Delta f_{LO}$, we found a 3.5% reduction in the measured acceleration but a 3.8% increase for $\Delta f_{SIG} < \Delta f_{LO}$.

We then make use of the magnified rapid passage effects in the frequency domain to remove these residual distortions in the time domain data. Conceptually, this processing is analogous to the lens system shown in Fig. 2d. Because we have the magnified response from the resonance, we can add in a larger focal length lens, or smaller quadratic phase profile, prior to the inverse Fourier transform to refocus and thereby remove the residual rapid passage effects in the time domain. The corrected time domain response is then defined as:

$$RF_n(t) = FT^{-1}\left\{RF_n(\omega)\exp\left(i\Phi_{err}(\omega) \pm i\frac{\Phi_{RF}(\omega^2)}{m}\right)\right\} \quad (8)$$

where $\Phi_{err}$ corrects for residual phase drift by eliminating the linear phase term between the $RF_{RES}$ and $RF_{BKG}$ in the frequency domain. The quadratic phase term in Eq. 8 that acts as the larger focal length lens in Fig. 2d is defined by the magnification in Eq. 6. The frequency domain quadratic term, $\Phi_{RF}$, in Eq. 8 is inversely related to the time domain quadratic phase term, $\phi_{RF}$, illustrated in Fig. 2b (and shown in Eq. 5). An example of the nearly distortion-free time domain data is shown in Fig. 2e. The horizontal axis for of these spectra is converted from time to frequency using the applied SIG chirp rate prior to the line shape fitting.

The resulting, distortion-free cavity mode spectra were fit using a Fano lineshape[18], producing a time trace of the cavity displacement. The fit uncertainty of the cavity position was typically 1 MHz, corresponding to a displacement uncertainty of 1 pm. These displacement data can then be inverted to yield the measured acceleration[14, 19]. We note that this inversion is only possible because of the large separation of the mechanical modes in our optomechanical accelerometer, with no other mechanical modes below 100 kHz. Further, the only parameters required in the conversion from frequency displacement to length displacement and then to acceleration are: the laser frequency, the cavity free-spectral range the mechanical resonance frequency and the mechanical quality factor[14]. Critically, each of these parameters can be readily experimentally measured and, as a result, the optomechanical accelerometer can be employed to measure SI-traceable acceleration without an external calibration.

To characterize the noise performance of the dual comb spectrometer, we placed the optomechanical accelerometer on an optical table surrounded by acoustic shielding. The resulting Allan deviation[20, 21] of the measured cavity acceleration noise can be found in the left panel of Fig. 3. The Allan deviation minimum occurs at 1930 μs, corresponding to $3\times10^{-3}$ m/s². From the initial point on this Allan deviation and the measurement rate of 100 kHz, we can calculate the displacement and acceleration noise floors as 2 fm Hz$^{-1/2}$ and $7\times10^{-5}$ m/s² Hz$^{-1/2}$ (8 μg Hz$^{-1/2}$, based on 1 g = 9.80665 m/s²), respectively. This displacement noise floor is a factor of 4 lower than our previous results[11] despite utilizing only 30% of the optical power incident on the cavity that was previously employed. This enhanced sensitivity can be largely attributed to the tighter cavity mode fit convergence that resulted from the phase corrections discussed above.

To assess the dynamic range of the dual comb spectrometer, we utilized the electromechanical shaker table in open-loop mode. This allowed for higher induced accelerations than were possible when the back-to-back reference accelerometer was utilized. As can be seen in the right panel of Fig. 3, we were able to measure accelerations as large as 236 m/s$^2$ (24 g). At these large accelerations, the cavity mode is traveling 20 GHz which is nearly the entire 22 GHz width of the optical frequency combs. We did not observe any significant deviation from linearity in the measured acceleration even over this extended range. From the largest measured displacement and the minimum of the Allan deviation, we can calculate the system's dynamic range to be $7.6 \times 10^5$ which we believe is the highest dynamic range demonstrated with a microfabricated cavity optomechanical sensor.

To assess the accuracy of the optomechanical accelerometer with the optical frequency comb readout, we also operated the electromechanical shaker in closed loop fashion using the reference accelerometer. Measurements were made across two days in Boulder, Colorado (referred to as Day 1 and Day 2) using the dual comb spectrometer shown in Fig. 1b and a single day roughly a month later in Gaithersburg, Maryland with a self-heterodyne single-comb spectrometer which has been previously described[12] (referred to as Day 3). These two locations offered considerably different temperatures, pressures, humidities, and elevations, thus allowing us to assess possible systematic effects from these experimental conditions. In addition, the LO chirp parameters were changed between Day 1 and Day 2, as shown in Fig. 2 (see also Figs. S3 and S4 in the Supplemental Material). Finally, the reduced comb bandwidth (2.2 GHz) of the self-heterodyne spectrometer leads to no observable rapid passage effects in the frequency domain (or any need for phase correction). As a result, these three days of measurements provided a means to assess potential systematic uncertainties, including those from the data processing and phase correction.

The results of the three sets of measurements made at a shaker drive frequency of 100 Hz can be found in the left panel of Fig. 4. We note that the measurements made across all three days are in good agreement with the reference accelerometer at the 2σ level. In addition, we see 1σ level agreement for all measurements above 5 m/s$^2$. Further, the optomechanical accelerometer exhibits far lower random (Type A) uncertainty than the reference accelerometer's reported uncertainty for all but the very lowest accelerations. We emphasize that unlike traditional piezoelectric or micro-electromechanical (MEMS) accelerometers, the optomechanical accelerometer does not require any form of external calibration, thus allowing it to serve as a high accuracy intrinsic standard[14].

As shown in the right panel of Fig. 4, the agreement with the reference accelerometer was seen over a wide range of frequencies below 2 kHz. However, above 2 kHz, we observed large deviations between the two accelerometers. These deviations, which are not unexpected, are likely due to mechanical modes in the stacked structure consisting of the shaker armature, reference accelerometer, and optomechanical accelerometer. We are currently designing alternate packaging of the accelerometer to reduce these effects as well as employing more advanced shaker tables and acceleration references.

A secondary advantage of the comb spectrometer over a laser locking approach is that information on the entire device line shape is available, not just the center position. Figure 5 illustrates how the line shape can be used to extract additional information such as optical coupling and finesse. For this data, the accelerometer was driven at 4 kHz at three different acceleration amplitudes. At this frequency, the adhesive used to assemble the accelerometer is no longer rigid. This leads to a reduction in the fitted cavity mode area (optical coupling efficiency) and cavity mode width (optical cavity finesse) at the portions of the sinusoidal signal corresponding to zero displacement (maximal velocity) for the largest drive levels. We note that this type of data could be used to evaluate various packaging methods and mechanically stable operating parameters.

The dual optical frequency spectrometer described here has allowed us to achieve an unprecedented combination of dynamic range, measurement speed, sensitivity, and accuracy for a microfabricated optomechanical sensor. In addition, we have shown that the magnification from the differential chirp down-conversion can be used to remove residual line shape distortions. The physical understanding and data processing control of the temporal dynamics of the optical system led to significantly improved sensing speed, sensitivity, and dynamic range. We also note that these approaches are also amenable to systems using lower cost chirp generation methods such as direct digital synthesis[22] and frequency multiplication. Beyond accelerometry, we envision impactful applications for this approach to a wide range of other optomechanical sensing targets including dynamic pressure, inertial navigation, and ultrasound. Finally, we note that these electro-optic comb methods are ideally suited to sensor networks in which a single pair of optical frequency combs could readily readout an array of optomechanical devices.

**Supplemental Materials**

The associated Supplemental Materials give further details on the modeling approach and methods.

**Acknowledgements**

Portions of the work described here were performed in the NIST Nanofab. This research was partially supported by the NIST-on-a-Chip program.

**Conflict of Interest Statement**

The authors have no conflicts to disclose.

**Author Contributions**

Conceptualization: DAL, JRS, BJR, TWL, DFP, JJG

Investigation: DAL, JRS, BJR, YB, FZ, SMB, DFP

Funding acquistion: DAL, TWL, DFP, JJG

Writing-original draft: DAL, JRS, BJR, DFP, JGG

Writing- review and editing: DAL, JRS, BJR, SMB, TWL, DFP, JJG

## Data Availability Statement

The data supporting this paper and the supplemental material will be made available at a DOI hosted by the National Institute of Standards and Technology.

## Figures

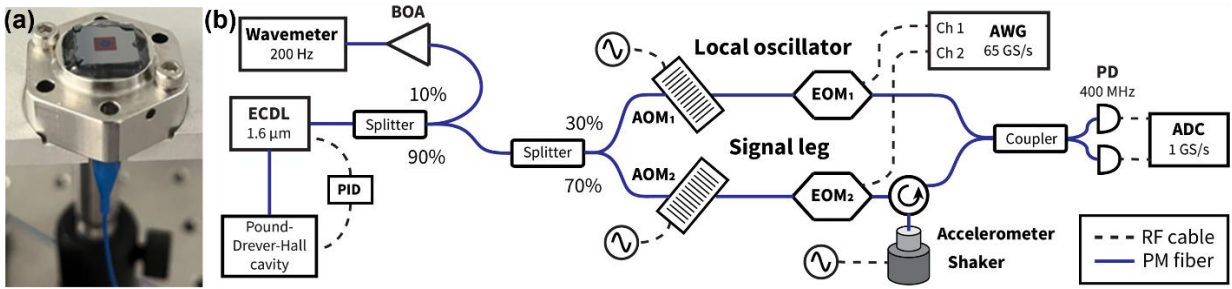

Figure 1. (a) Fiber coupled optomechanical accelerometer mounted into a stainless-steel enclosure. The enclosure also includes a cover (not shown) to protect the photonic chips and enable mounting on an electromechanical shaker table. (b) Instrumental schematic of the electro-optic dual comb spectrometer. Two electro-optic frequency combs are produced, with one probing the accelerometer (Signal, SIG) and the other serving as a local oscillator. The abbreviations are external-cavity diode laser (ECDL), booster optical amplifier (BOA), proportional-integral-derivative servo (PID), acousto-optic modulator (AOM), electro-optic phase modulator (EOM), arbitrary waveform generator (AWG), photodetector (PD) and analog-to-digital converter (ADC).

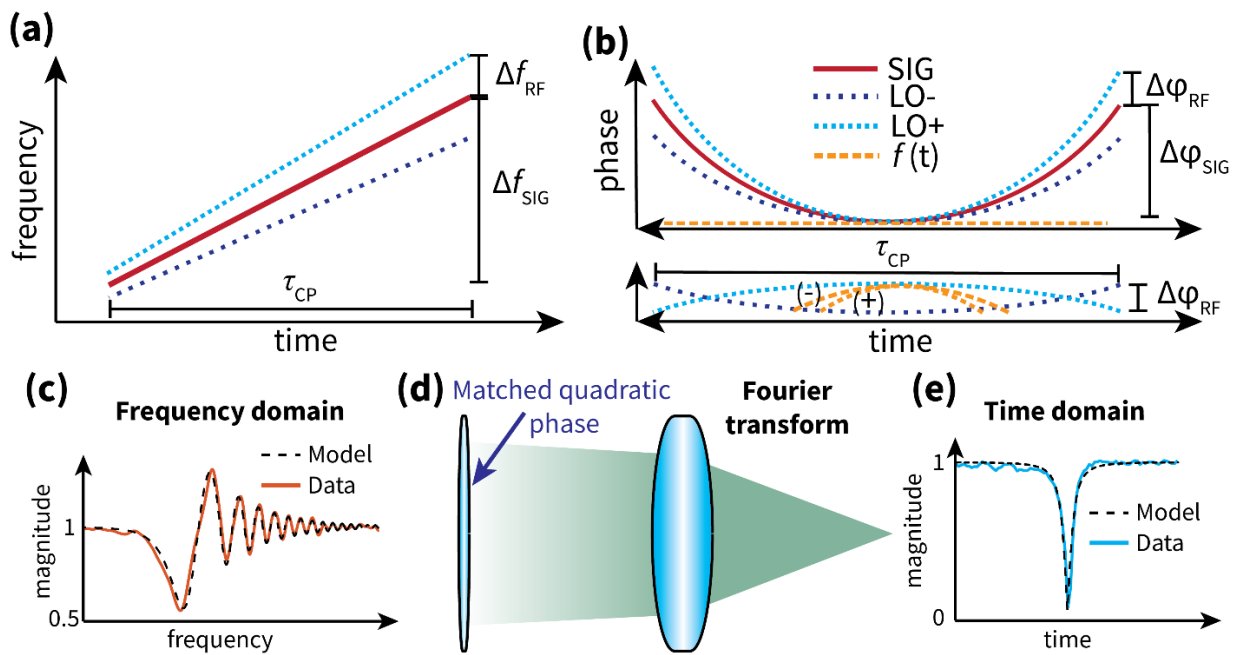

Figure 2. a) The signal chirp (SIG) spans the bandwidth $\Delta f_{SIG}$ in the duration, $\tau_{CP}$, and is down-converted using a local oscillator (LO) that is either chirped more (i.e., $\Delta f_{SIG} < \Delta f_{LO}$, LO+) in dashed light blue, or less (i.e., $\Delta f_{SIG} > \Delta f_{LO}$, LO-), in dotted purple than the signal chirp shown in solid red. b) The quadratic phases of the SIG and LO chirps as a function of time. These phases are subtracted to map them into the radiofrequency (RF) domain (lower panel). The phase response from the resonance, shown in long dashed yellow, is magnified by the quadratic phase of the LO. c) The frequency domain data in red together with the model data overlaid (dashed black) showing the magnified rapid passage behavior. Details of this model can be found in the Supplemental Material. d) The Fourier transform in time is analogous to a spatial lens, where adding an additional lens serves to remove (reshape) the residual rapid passage effects. e) The resulting distortion-free cavity mode spectrum following inverse Fourier transformation.

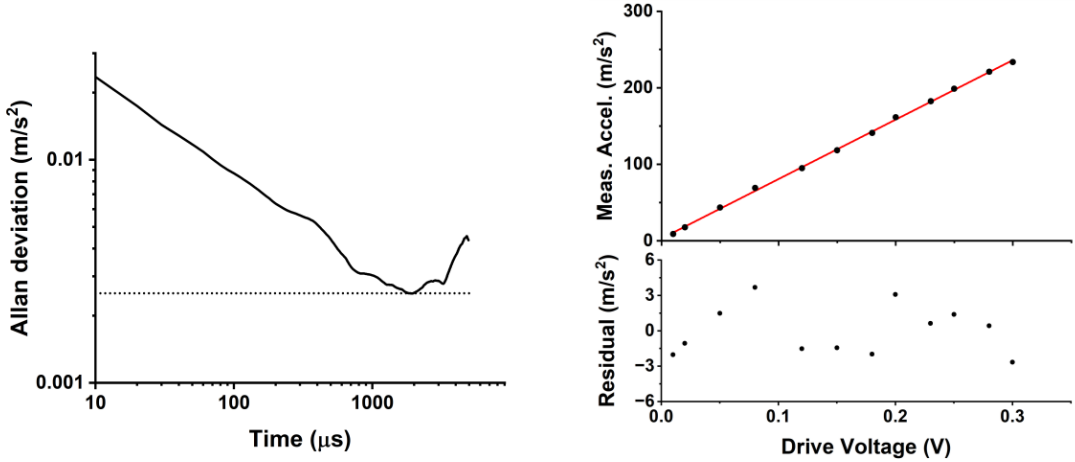

Figure 3. (Left panel) Overlapped Allan deviation[20, 21] of the amplitude of the acceleration measured with the optomechanical accelerometer when it was placed on its side in an acoustic enclosure. The conversion from displacement to acceleration was made assuming that the noise was occurring near DC. (Right panel) Acceleration measured by the optomechanical accelerometer as a function of the root-mean-square voltage applied directly to the electromechanical shaker table. The corresponding uncertainties are also plotted but are smaller than the shown data points. The red line gives a linear fit to the measured data. The accelerometer and shaker combination appear linear over a wide range of accelerations, even at accelerations as large as 236 m/s$^2$ (corresponding to 24 g where 1 g = 9.80665 m/s$^2$).

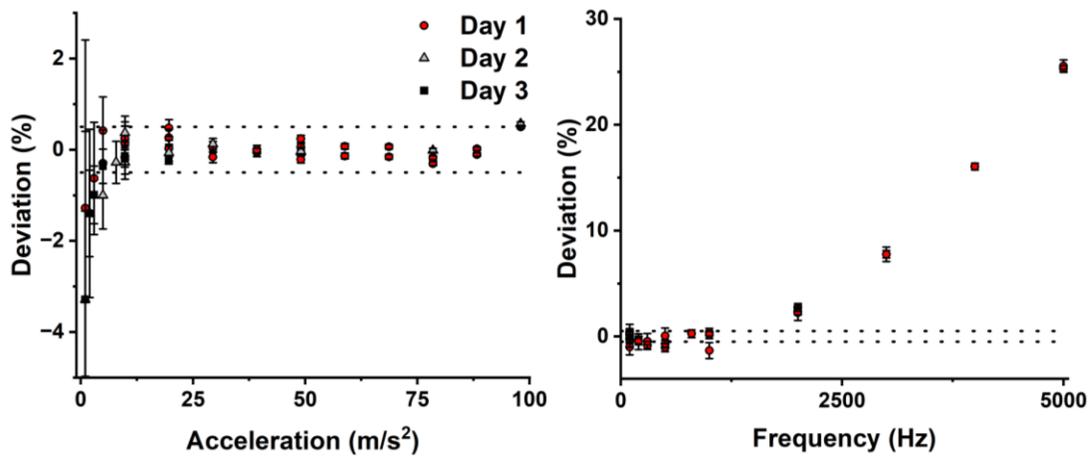

Figure 4. (Left panel) Deviations between the accelerations measured by the optomechanical accelerometer and the back-to-back (stacked) reference accelerometer as a function of applied acceleration at 100 Hz. The uncertainties on the data points were based on the standard deviations of the data described in the left panel of Fig. 3, while the dotted

lines illustrate the ±0.5 % (k=1) uncertainty associated with the reference accelerometer's calibration. Measurements are shown from three separate measurement runs. Day 1 and Day 2 correspond to measurements taken with the dual comb spectrometer in which the direction of the chirps driving the electro-optic modulators were reversed (leading to opposite rapid passage effects). Day 3 data were recorded in a separate location using a self-heterodyne comb spectrometer which does not exhibit rapid passage effects due to its more limited comb bandwidth (2.2 GHz) [12]. (Right panel) Measured deviations between the optomechanical accelerometer and the back-to-back reference accelerometer for accelerations between 5 m/s$^2$ and 10 m/s$^2$ as a function of excitation frequency. These measurements were recorded across Day 1 and 2 with the dual comb spectrometer.

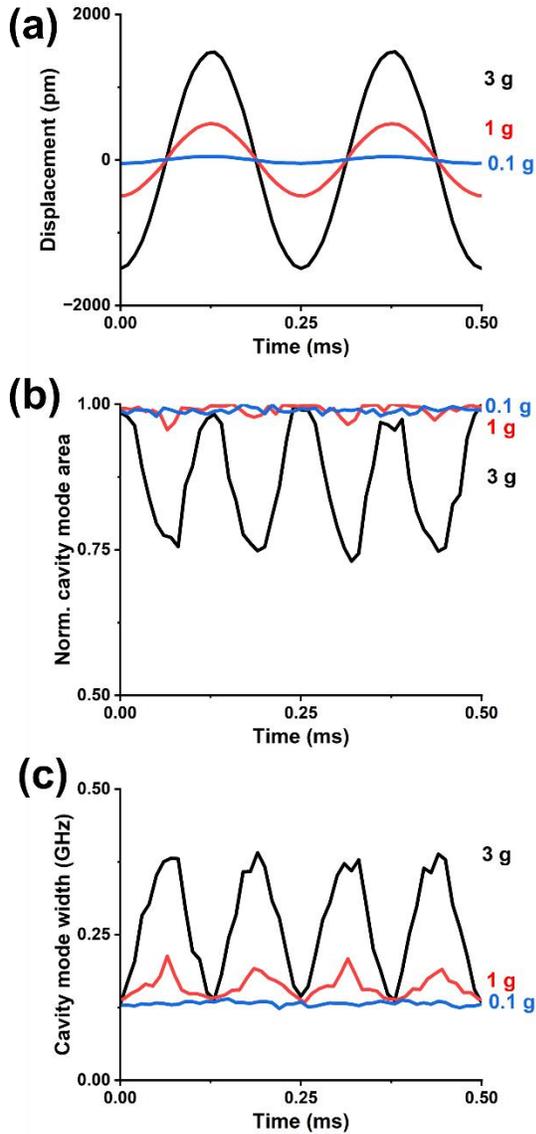

Figure 5. Dynamics of the optical cavity line shape during a 4 kHz excitation at nominal accelerations of 3 g (black), 1 g (red), and 0.1 g (blue), where 1 g = 9.80665 m/s$^2$. The line shape center, amplitude, and width are modulated, especially at higher drive levels, leading to variations in the a) cavity displacement, b) mode area and c) mode width.

11